# HSBC until 1950: From its colonial cradle past the World Wars

by Christopher Mantzaris[*], Prof. Dr Ajda Fošner[*].

**Abstract:**

Europe's largest bank by assets as of 2025 started out in the 1860s in one of Europe's colonies: The Hongkong and Shanghai Banking Co (HSBC). Multiple wars forced Qing China and later the young Republic of China into a series of unequal treaties, one of which was the forced legalisation of the opium trade from parts of the British Empire into China, another was opening several cities, including Shanghai, for trade and granting extensive civil, property and business rights to non-residents and yet another was the annexation of Hong Kong by the United Kingdom. These are the conditions that created HSBC and in which it thrived, including from opium-related profits. During periods of relative calm, the bank grew geographically and made profits –whether in moral or unethical, whether in legal or unlawful ways– which helped HSBC weather the storms of civil and world wars. Other aspects contributed to HSBC's survival and success, such as its global nature, which allowed it to diversify and shift away from regions when danger emerges there and find shelter in safer havens. Yet the resilient survival abilities and the financial success of HSBC until 1950 should not distract from the fact that in addition to its tainted cradle and early profits from facilitating the poisoning of a whole society, its human resource system was also discrimination based, attempting to divide the one human race into different groups – in spite of such practises being opposed to medical and biological facts. What is particularly interesting to see is that not only has HSBC yet to apologise for the early drug-related blood money it made: it even fails to mention its colonial, drug profits tainted past on any of its many history-themed pages sighted. Later sections contain possible reasons for HSBC's resilience and success, particularly interesting for entrepreneurs.

# Introduction

One of the largest contemporary banks in the world and the largest in Europe, initially named *The Hongkong and Shanghai Banking Co*[1] or *Hongkong and Shanghae Banking Co*[2], then –after incorporation– *The Hongkong and Shanghai Banking Corporation*[3,4,5] –since 1991 and as of 2025 anyway: *HSBC Holdings plc*[6,7] (HSBC)– dates its roots back to the year 1864 or 1865 in British occupied Hong Kong and Shanghai.[8] 160 years later, as of 2025, it is the seventh largest bank in the world by assets and the largest in Europe,[9] which stand at roughly 3 Trillion (Tr) United States (US) dollars.[10] In between, many world-shaping events occurred: China and the United Kingdom shifted away from a hereditary, absolute monarchy – with differing degrees of success. Civil wars and regime changes shock the region and two world wars the globe. Communist-labelled authoritarianism devastated the Chinese economy and society, resulting in massive amounts of killings and destruction of anyone and anything that was perceived as non-communist or associated with capitalism, such as entrepreneurs, land owners or even scholars – In particular during and after a Civil War, intermitted by World War Two, and during the Cultural Revolution.[11,12] Yet, what is more emblematic of capitalism than the global banker and what can better fit the image of communism's adversary than a foreign, debt-collecting bank?

---

[*] Address and affiliation: University of Primorska, Koper; CM's email: 68223065[at]student.upr.si.
1. history.hsbc.com/collections/global-archives/the-hongkong-and-shanghai-banking-corporation-1/board-minutes/extract-from-the-first-minutes-of-the-board-of-the-hongkong-and-shanghai-banking-corporation-6-aug-1864/1702323
2. scmp.com/article/689439/bank-was-built-shipping-and-opium-trade
3. oelawhk.lib.hku.hk/items/show/780
4. web.archive.org/web/20130131202233/http://www.hsbc.com/about-hsbc/history/hsbc-s-history
5. web.archive.org/web/20101204041118/http://www.hsbc.com/1/2/about/history/1865-1899
6. web.archive.org/web/20130131202233/http://www.hsbc.com/about-hsbc/history/hsbc-s-history
7. web.archive.org/web/20101204041118/http://www.hsbc.com/1/2/about/history/1865-1899
8. history.hsbc.com/collections/global-archives/the-hongkong-and-shanghai-banking-corporation-1/board-minutes/extract-from-the-first-minutes-of-the-board-of-the-hongkong-and-shanghai-banking-corporation-6-aug-1864/1702323

How did HSBC not only survive such a tumultuous history and seemingly hostile environment, but how could it apparently even thrive tremendously despite –or because of– all that? Why, in what background and under what circumstances was this bank founded in the first place? How did HSBC suffer from, was involved in or profited off of injustices and how did it apologise and amend, if any?
Also, what lessons can entrepreneurs draw from this case study, to mimic HSBC's long-standing survival and reach prosperity in spite of the adverse conditions and destructive chaos it faced? This research project collects data to answer these questions.

## The Background that birthed HSBC: The start of British imperialism in China, military subjugation, unequal treaties and illegal opium trade

In 1711, the British East India Company set up their base in Canton –locally and more modernly known as Guangzhou– to trade Chinese goods, like silk ceramics and tea.[13] Later the same century, >60% of East India Company's purchases in China were tea.[14] These tea purchases were increasingly financed with Indian opium sold in China, even after the drug trade was fully illegalised by the Qing Dynasty in 1796.[15] A previous ban was already in place since 1729, which forbade Madak, a mixture of opium with tobacco.[16] This was the most common form of opium, as it opium was otherwise hard to preserve at the time.[17] Neither the initial ban, nor the more comprehensive ban in 1796 deterred the British: merely the location changed from Guangzhou to the small, nearby *Lintin* or *Nei Lingding Island*, right in the bay bordering –and not far from–

---

9  https://web.archive.org/web/20251106111745/https://www.spglobal.com/market-intelligence/en/news-insights/articles/2025/4/the-worlds-largest-banks-by-assets-2025-88424232
10 https://web.archive.org/web/20251106111745/https://www.spglobal.com/market-intelligence/en/news-insights/articles/2025/4/the-worlds-largest-banks-by-assets-2025-88424232
11 https://perma.cc/N68Q-LRNE
12 https://perma.cc/8UYX-S29S
13 web.archive.org/web/20170802/howardscott.net/4/Shameen_A_Colonial_Heritage/Files/Journal.html
14 web.archive.org/web/20170802/howardscott.net/4/Shameen_A_Colonial_Heritage/Files/Journal.html
15 web.archive.org/web/20170802/howardscott.net/4/Shameen_A_Colonial_Heritage/Files/Journal.html
16 Peter Ward Fay, *Opium War, 1840-1842: Barbarians in the Celestial Empire in the Early Part of the Nineteenth Century and the War by Which They Forced Her Gates,* Chapel Hill of North Carolina 1997, 73.
17 Charles C. Mann, *1493: Uncovering the New World Columbus Created,* New York City 2011, 123–163.

Guangzhou, Shenzhen, Macao and Hong Kong.[18] Yet, the small island in the Canton province was unlikely the only place of opium import, as the 6th Qing emperor, Daoguang, also mentioned in 1810 Fujian in addition to Canton as a place from which opium entered China.[19]

As the effects of the drug trade continued to worsen and even the Qing emperor's son died from opium, the Emperor ordered the public destruction of all opium subsequently seized from foreigners.[20] In addition, there were minor military hostilities over a local who was killed by the foreigners around Hong Kong (*Kowloon Incident*): The killers were subsequently shielded by the British East India Company from prosecution, resulting in a smaller skirmish between British and Qing personnel.[21] This was enough for the British empire to subjugate Qing China with their naval power in a war that started with the Kowloon Incident on 4 September 1839 and ended on 29 August 1842 with the Treaty of Nanking.[22] This First Opium War and its treaty were only the first in the subsequent series of wars and *unequal treaties* in which Western powers used alliances and military superiority to wrest more and more land and concessions from Qing China and later the young Republic of China.[23] The treaty ceded Hong Kong Island to Britain in perpetuity, gave foreigners property, residence and other extensive civil rights and forced foreign trade to also be allowed in five additional Chinese cities, namely Shanghai, Canton (Guangzhou), Ningbo, Fuzhou, and Xiamen.[24] One of the subsequent unequal treaties –after the Second Opium War, which concluded in 1860– was the Convention of Peking of 24 October 1860 and its supplements, which stipulations included:
— The Hong Kong colony becomes larger, by permanently including adjacent Kowloon (spelled as *Cowloon),*
— Other territorial losses to Russia,
— Tianjin (in the treaty: *Tientsin* or *Tien-tsin*) also becomes a trade port,
— Legalisation of the opium trade,
— Freedom of religion in China and full civil rights for foreigners/Christians,
— Indentured Chinese can be carried to the Americas with British ships,
— Monetary reparations.[25]

All this allowed opium sales to China to continue their exponential growth:

---

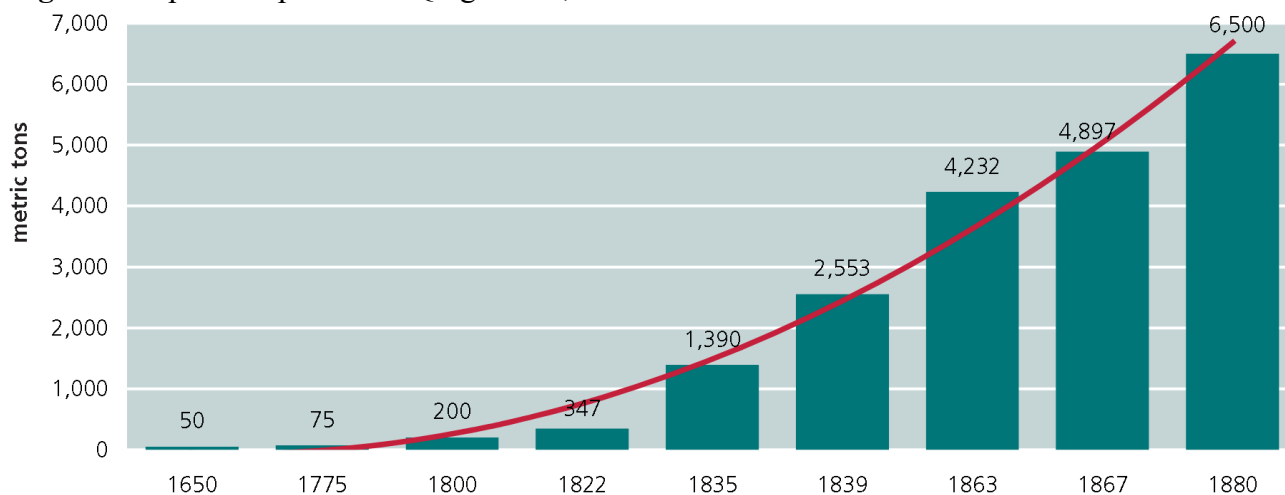

**Figure 1:** Opium imports into Qing China, from 1650 to 1880

*Sources:*
Graph taken form: United Nations, *World Drug Report 2008*, Vienna 2008, 175 (unodc.org/documents/wdr/WDR_2008/WDR2008_100years_drug_control_origins.pdf).
Cited there: Thomas D. Reins, *The Opium Suppression Movement in China*, Modern Asian Studies, Vol. 25 No. 1, 1991;
Michael Greenberg, *British Trade and the Opening of China*, 1800-1842, Cambridge (England) 1947;
Fred W. McCoy, *The Politics of Heroin*, New York 1991.

It was this environment that necessitated for the British a bank on site, to facilitate the sale of Indian opium in China and the use of these funds to purchase Chinese goods, especially tea, silk and china – for further sale in Europe.[26]

## The people and characters who started HSBC

In 1834, Thomas Sutherland was born in Aberdeen, Scotland, where he –due to the early death of his father– was soon raised by his mother's wealthy parents.[27] He attended a grammar school, which was headmastered by[28] the esteemed Dr James Melvin LLD (1795–1853).[29] Subsequently –between the ages of 14 and 15– he studied at Aberdeen University.[30] After that early and short year in tertiary education, he first worked as an office worker for a merchant in Aberdeen, then –before turning 18– he applied with success through a friend of his mother and aunt at the Peninsular and Oriental Steam Navigation Company (P & O), a shipping company[31] which name still exists today, despite starting out in 1822 and being formally founded on August 22 1837.[32] Sutherland quickly rose in ranks, as he was promoted within a few years first to Bombay, then to Hong Kong, where he become superintendent of the China and Japan affairs, with much freedom to run and expand the businesses there how he sees fit.[33] In 1866, the year after Sutherland started HSBC, he founded the

---

26   web.archive.org/web/20081211/hoovers.com/globaluk/sample/co/history.xhtml?ID=ffffcrksftxxchrftc
27   electricscotland.com/history/other/sutherland_thomas.htm
28   web.archive.org/web/20220223/abdn.ac.uk/special-collections/cld/30
29   electricscotland.com/history/other/sutherland_thomas.htm
30   electricscotland.com/history/other/sutherland_thomas.htm
31   cruisecritic.co.uk/articles/po-cruises-history-from-the-1800s-to-the-present-day
32   electricscotland.com/history/other/sutherland_thomas.htm
33   electricscotland.com/history/other/sutherland_thomas.htm

*Hong Kong Kowloon and Whampoa Dock Company*[34], which is as of 2025 is part of the *CK Hutchison Holdings Limited* conglomerate – listed on the Stock Exchange of Hong Kong with the Symbol "1", now one of the 500 largest companies in the world.[35] After HSBC opened for business in 1865 –the first board meeting was held in 1864[36] and the construction of the building for it started also in 1864[37]–, the then 31 year old Sutherland seemed to never have made HSBC his personal priority, or at least not his main life-project: He may have spent the first 2 to 3 years as vice president and principal working director at HSBC, though then seemed to have mostly turned his attention to other issues.[38,39,40] If there was anything like a primary focus of his –he was a man with many founded companies and many missions and tasks he engaged in– it was most likely P & O.[41] It therefore seems as if it was more the combination of reasons, namely that HSBC was started rather at the right time, at the right place, by the right –a privileged and also talented– man, but then gained quickly its own momentum – as opposed to HSBC having started as a personal project, which was built by the relentless endurance and the blood and sweat of highly motivated, individual empire builders. If there was an exception in the earlier years of HSBC –any single, nameable figure who stood out during HSBC's start or shortly thereafter– however, it was most likely the Carrigallen, Leitrim County[42,43], Ireland born Thomas Jackson, who led the bank in its most senior position of chief manager, for the period from 1876 –when he was 35[44]– until 1902. Yet not from April 1886 to September 1887, not in 1889, nor between 1891 and 1893, when he relocated to the United Kingdom to instead run HSBC's London branch.[45,46] Jackson worked in the bank since the first year in 1865 and became to be called by some "Great Architect" and "Lucky Jackson", likely due to his smart choices.[47]

Noteworthily, albeit Sutherland and Jackson being the figures that are as of 2025 most presently and loudly mentioned, Sutherland founded HSBC with other people, including with Thomas Dent, founder of the drug smuggling company *Dent & Co* – which underlines and gives further evidence to the importance of the at first illegal but by then at least unethical drug-related money that HSBC made in its early years.[48]

---

# The start and early years of HSBC: 1864-1900

The immense, strongly growing trade with China, most notably in the new British colony Hong Kong with Kowloon and the city now open for trade, Shanghai, Sutherland saw the potential for a trade bank.[49] A trade bank's business includes guaranteeing sellers and buyers, exporters and importers, that debts will be paid –such as by paying for debts themselves if their client fails to do so–; or help them finance their transactions – for example by providing credit to an exporter to buy goods, until the new buyer pays for them.[50] Sutherland had trouble understanding why the locals nor any other merchant started a bank in the newly yet booming region, and instead relied on banks in England and India.[51] By the time Sutherland started the bank's operations, on 3 March 1865, opium already encompassed 70 per cent of what the British Empire exported into China.[52] Already by April the same year, Sutherland opened a HSBC branch in Shanghai.[53,54] More expansions followed: in 1866 in Yokohama, Japan, but also Calcutta, India in 1867 as well as Saigon (Ho Chi Minh City), Vietnam by 1875, Manila, Philippines in 1875 and in November the same year in San Francisco: a total of seven countries by 1875[55,56,57]. Singapore followed in 1877, New York in 1880[58] and in 1888, HSBC established Thailand's very first bank at the old Belgian Consulate on Bangkok's Charoen Krung Road,[59] which even printed the country's first banknotes[60,61,62] It is quite interesting to note, that even as of 2020, between 40 to 50 per cent of Hong Kong's legal tender –Hong Kong dollar– notes are printed by HSBC.[63]

Rather quickly, still in its first decade, the bank also become involved in government businesses, such as the issuance of government debt for infrastructure project, as exemplified by HSBC's placing of China's first public government loan in 1874.[64]

The first years were not as successful, contrary to the 1880s, during which HSBC saw tremendous success under Jackson, when HSBC was the British government's bank of choice in the Orient.[65] By circa 1900, opium has largely receded as an import into China and was instead replaced largely by other goods[66], as also illustrated in below Figure 2:

---

**Figure 2:** Opium imports into Qing China, from 1880 to 1908

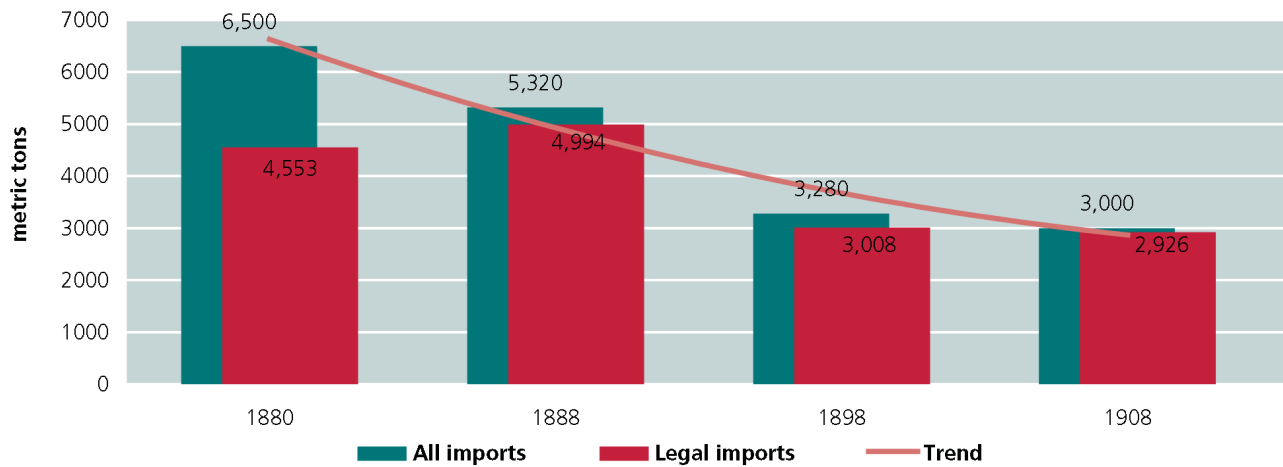

*Sources:*

Graph taken form: United Nations, *World Drug Report 2008*, Vienna 2008, 176 (unodc.org/documents/wdr/WDR_2008/WDR2008_100years_drug_control_origins.pdf).

Cited there: International Opium Commission, *Report of the International Opium Commission*, Shanghai: 1909, Vol. II, Reports of the Delegation, Memorandum on Opium in China, 46-47;

Observatoire Géopolitique des Drogues, *Atlas Mondial Des Drogues*, Paris 1996, 27;

Thomas D. Reins, *Reform, Nationalism and Internationalism, "The Opium Suppression Movement in China and the Anglo-American Influence, 1900-1908"*, 1991, 114.

Given the importance of the historical context in explaining why HSBC came into existence –the imperial, colonial reality obtained and enforced through violence– and considering what laid the profitable foundation from which HSBC grew into what it is now –the illegal opium trade, destructive for millions of individuals and families–, it is rather unfaithful of HSBC to mention neither the words "opium", "drugs", "narcotics" etc, nor the words "colonial" or "colony" etc anywhere on HSBC's multiple pages sighted between 2009 and 2025, where HSBC claims to inform readers about its own history.[67,68,69,70,71,72,73,74] To make matters worse, the United States site of HSBC even mentions proudly the products which trade HSBC helped finance in its early years, yet conveniently leaves out opium, which trade facilitation was almost certainly the company's core business during just that time[75,76]: "By 1875 HSBC [...] financed the export of tea and silk from China, cotton and jute from India, sugar from the Philippines and rice and silk from Vietnam."[77] The attempted ESG-washing hardly distracts from the lack of actual ethical and social responsibility.[78]

---

67  web.archive.org/web/20090912003609/http://www.hsbc.com/1/2/about/history/1865-1899
68  web.archive.org/web/20130131202233/http://www.hsbc.com/about-hsbc/history/hsbc-s-history
69  web.archive.org/web/20170428050017/http://www.hsbc.com/about-hsbc/company-history/hsbc-history
70  web.archive.org/web/20210503055136/https://www.hsbc.com/who-we-are/our-history
71  web.archive.org/web/20210428143814/https://www.hsbc.com/who-we-are/our-history/history-timeline
72  web.archive.org/web/202501/history.hsbc.com/exhibitions/a-bank-at-war
73  web.archive.org/web/202501/hsbc.com/who-we-are/purpose-values-and-strategy/our-history/history-timeline
74  web.archive.org/web/202501/about.hsbc.co.uk/history-timeline
75  www.grandprix.com/gpe/spon-022.html
76  scmp.com/article/689439/bank-was-built-shipping-and-opium-trade
77  web.archive.org/web/202501/about.us.hsbc.com
78  sciencedirect.com/science/article/abs/pii/S0261560624000305

# HSBC in the new century and a world at war: 1900-1920

Had HSBC by 1875 operations in 7 countries, by 1900, that number had more than doubled to 16.[79] Early into the 20th century, HSBC become more and more involved in assisting governments issuing their debt –in particular in China– so that governments can use these funds to conduct modernisation and infrastructure investments, such as investing in the new technology of the previous century, which gained more and more popularity: Rail and tram transportation.[80] Hong Kong's tramway system, for example, was approved on 29 August 1901 by England, construction of the rails followed in 1903 and operations started on 30 July 1904 – with 26 tramcars built in the United Kingdom, transported to Hong Kong in pieces, and finally reassembled there.[81] Two years before, in 1902, the first Electric Tramway in East Asia opened in the Kanagawa Prefecture.[82] After Kanagawa and Hong Kong, Dalian, in Qing China, started tram operations in 1909,[83] followed by 17 railway systems in Japan between 1910 and 1933.[84,85,86] This all happened in addition to heavy rail construction or expansion in many of these and other places. The bank used this period of stability, growth, investment and globalisation – and paired it well with the ever increasing trust the United Kingdom, its crown and government as well as other governments all around the global, especially in Asia, placed in the bank.

Concurrently with the expansion of its business with governments, HSBC also grew its branch network: In the first decade of the 20th century, HSBC grew its branches from a proud 300 to an incredible 650.[87]

HSBC, still headquartered and incorporated in Hong Kong and mainly focused on East Asia, was less involved in World War I, despite being a British bank – The more United Kingdom focused Midland Bank, which merged with HSBC in 1992, was much more affected by the First World War.[88] Yet, even at HSBC thousands of staff members likely fought in World War I, which resulted in a higher percentage of women in the work force.[89] From HSBC's London office alone, 169 HSBC employees joined the British military for World War I.[90] Out of these 169, 24 per cent –40 people– died in or never returned from the war.[91]

One of the reasons why HSBC was likely less affected by the First World War than other banks in the British empire, might have been due to the fact all relevant East Asian jurisdictions were in the

---

Entente and hence Allied Powers[92], which naturally should have resulted in no hostilities, death or destruction –and less business disruptions– where HSBC was most present and headquartered: in Hong Kong and the surrounding area – in stark contrast to Europe.

## The 1920s, civil war and the Great Depression: HSBC 1920-1937

After the First World War, HSBC further expanded in East Asia, including by opening further branches in Bangkok in 1921, Manila in 1922 and in 1923 Shanghai,[93] spurred by thriving tin and rubber trade in the region.[94] Yet, this renaissance of globalism, deeper world-wide economic cooperation and prospertiy was brief and mostly limited from circa 1919 to 1927 – also for the bank.

In the Great Depression of 1929 and the years following it, HSBC decided to reduce or totally cancel bonuses for its staff and lower –yet not scrap– dividends for its shareholders.[95] In this period, as HSBC's reserves were shrinking, it also decided against expansion and focused on pure survival instead.[96] This was exacerbated by the Civil War between the nationalist Kuomintang –which formed the official government of the Republic of China– and the Communist Party of China.[97] This civil war raged on from 1 August 1927, which was only paused shortly after the invasion of the Republic of China by the Empire of Japan.[98] 1 August 1937 was the true start of the largest global conflict and disaster humanity as ever seen: the Second World War; which was expanded two years later, where it reached Europe, when Hitler's National-socialist –better: fascist– Germany invaded Poland on 1 September 1939.

## HSBC during and after the Second Civil and World War: 1937-1950

The full scale invasion of the Republic of China by the Empire of Japan began not far from Beijing, with the Marco Polo Bridge incident on 7 July 1937.[99] In December 1936 or early 1937, in anticipation of the Empire of Japan's aggression, most of the government of the Republic of China, led by the Kuomintang, entered into an uneasy peace with the Communist Party of China, to focus on defending against the Empire of Japan's invasion.[100] Japan's surrender, which marked the end of World War Two, only occurred on 2 September 1945 globally, yet separately for the Republic of China only a week later, on 9 September 1945.[101] Yet, this did not end hostilities, destruction, death and war in China, as the Chinese Civil War –merely paused to a certain degree by the Empire of

---

92  britannica.com/topic/Allied-powers-World-War-I
93  web.archive.org/web/2010120/hsbc.com/1/2/about/history/1900-1945
94  hsbc.com/who-we-are/purpose-values-and-strategy/our-history/history-timeline
95  hsbc.com/who-we-are/purpose-values-and-strategy/our-history/history-timeline
96  hsbc.com/who-we-are/purpose-values-and-strategy/our-history/history-timeline
97  britannica.com/event/Second-Sino-Japanese-War
98  britannica.com/event/Second-Sino-Japanese-War
99  britannica.com/event/Second-Sino-Japanese-War
100 britannica.com/event/Second-Sino-Japanese-War
101 britannica.com/event/World-War-II -> Chapter: "The Japanese surrender".

Japan's aggression– enraged anew after Japan's surrender in September 1945 and continues somewhat until today: between the authoritarian People's Republic of China in Beijing and the democratic Republic of China in Taipei. Yet the war in the narrower sense ended mostly with the Communist Party of China's conquest of the Hainan, Zhoushan and Wanshan islands, which concluded between April and August 1950.[102]

In this regional, sociological and geopolitical background –and quite violent, destructive times–, HSBC arguably struggled and suffered more than it ever did during its history, before or since – the only potential contender, at least for the bank's financial affairs, might be the 2008 financial crisis. Most of the bank's East Asia staff become prisoners of war, due to their affiliation with the British Empire, which of course was opposed, as an Ally, to the Axis Power: The Empire of Japan.[103] HSBC's most senior staff, the chief manager, Sir Vandeleur Grayburn, as well as the successor he designated, died while the were being held captive by forces of the Empire of Japan.[104] In 1941[105] or 1943[106], the HSBC branch in London was formally declared the new headquarters – after pressure from British law makers to do so. Also in 1941[107] or 1943[108], Westminster likewise insisted that the bank's London Advisory Committee should be HSBC's new board of directors and in consequence, that the committee's Arthur Morse become the chairman and new chief manager in personal union – which is what happened. The brank survived this period on its reserves.[109] In the midst of World War Two, on June 1944, the management in London did not recommend a dividend for the past year of 1943, while it already suggested to pay again cash to the bank's shareholders by April 1947, for the fiscal –and calendar– year of 1946: The first post-war year.[110,111] By then, HSBC also already moved its headquarters back to Hong Kong.[112] There, in post World War Two Hong Kong, HSBC had a crucial role in the astonishing rebuilding and economic growth that occurred in British Hong Kong, by providing its services, networks and expertise – but also by attracting talent and newcomers to Hong Kong.[113,114] In the 1949 founded, communist People's Republic of China, led by the dictator Mao Zedong, however, HSBC had to start closing branches as early as 1949.[115]

---

102 Roderick MacFarquhar et al., *The Cambridge History of China*, Cambridge (England) 1991, 820.
103 web.archive.org/web/2010120/hsbc.com/1/2/about/history/1900-1945
104 web.archive.org/web/2010120/hsbc.com/1/2/about/history/1900-1945
105 www.grandprix.com/gpe/spon-022.html
106 web.archive.org/web/2010120/hsbc.com/1/2/about/history/1900-1945
107 www.grandprix.com/gpe/spon-022.html
108 web.archive.org/web/2010120/hsbc.com/1/2/about/history/1900-1945
109 web.archive.org/web/20130131202233/http://www.hsbc.com/about-hsbc/history/hsbc-s-history
110 history.hsbc.com/collections/global-archives/the-hongkong-and-shanghai-banking-corporation-1/annual-report-and-accounts/annual-report-and-accounts-of-the-hongkong-and-shanghai-banking-corporation-for-the-year-ending-31-dec-1943/1858647
111 history.hsbc.com/collections/global-archives/the-hongkong-and-shanghai-banking-corporation-1/annual-report-and-accounts/annual-report-and-accounts-of-the-hongkong-and-shanghai-banking-corporation-for-the-year-ending-31-dec-1946/1858558
112 history.hsbc.com/collections/global-archives/the-hongkong-and-shanghai-banking-corporation-1/annual-report-and-accounts/annual-report-and-accounts-of-the-hongkong-and-shanghai-banking-corporation-for-the-year-ending-31-dec-1946/1858558
113 http://web.archive.org/web/20130131202233/http://www.hsbc.com/about-hsbc/history/hsbc-s-history
114 http://web.archive.org/web/20101204064823/http://www.hsbc.com/1/2/about/history/1946-1979
115 http://web.archive.org/web/20101204064823/http://www.hsbc.com/1/2/about/history/1946-1979

# Conclusions

*Ethics optional*: The case study of HSBC suggests ethics to be not required, even when an entity aims for long-term success.

*Strong foundation*: HSBC was well capitalised at its foundation and run by founders with previous and parallel business success as well as wide and deep networks. It always chose the best locations, highest quality buildings and so on, even since its earliest branches. This gave it a strong base.

*Early momentum:* Building up on that strong base was the early gathering of momentum. HSBC expanded aggressively when it could and it started growing branches and business right from the start. The bank seemed to have lived the business mantra "Grow or die"[116] long before it was established as such.

*Reserves and rainy day funds*: HSBC had countless rainy days, not least wars and communism. It looks like it also managed to survive the lean years because it grew and build reserves during the fat years.

*Diversification*: The firm diversified not only in terms of revenue streams, as it appeared unaffected by the drying up of the opium trade. Foremost the geographically diversification seemed to have allowed it to survive as an entity, even when its home and core markets were engulfed in destruction.

*Stressors*: The bank is no stranger to crises. Its experiences during them must have made it more aware of the random, chaotic, unforeseeable nature of reality, which likely contributed to HSBC persevering to this day.

---

116 https://search.worldcat.org/title/352245884